\documentclass[twocolumn,showpacs,prl,aps,superscriptaddress]{revtex4}

\usepackage{amsmath}
\usepackage{amssymb}
\usepackage{graphicx}
\usepackage{upgreek}
\usepackage{bm}

\newcommand{\vect}[1]{\mathbf{#1}}
\newcommand{\ten}[1]{\mbox{\textbf{
{\textsf{#1}}}}}

\newcommand{\veczero}{\mathbf{0}}
\newcommand{\tenszero}{\mbox{\textbf{\textsf{0}}}}
\newcommand{\sprod}{\!\cdot\!}

\newcommand{\vprod}{\!\times\!}

\newcommand{\trans}{\mathsf{T}}
\newcommand{\dual}{{\circledast}}
\newcommand{\dif}{\mathrm{d}}
\newcommand{\mi}{\mathrm{i}}
\newcommand{\me}{\mathrm{e}}
\newcommand{\GRe}{\mathcal{R}\mathrm{e}}
\newcommand{\GIm}{\mathcal{I}\mathrm{m}}

\begin{document}

\title{Macroscopic quantum electrodynamics and duality in non-local
and Onsager-violating media}

\author{Stefan Yoshi Buhmann}
\email{s.buhmann@imperial.ac.uk}
\author{David T. Butcher}
\author{Stefan Scheel}
\affiliation{Quantum Optics and Laser Science, Blackett Laboratory,
Imperial College London, Prince Consort Road, London SW7 2AZ, United
Kingdom}

\date{\today}

\begin{abstract}
We formulate macroscopic quantum electrodynamics for the most general
linear, absorbing media. In particular, Onsager reciprocity is not
assumed to hold. For media with a non-local response, the field
quantisation is based on the conductivity tensor and the Green tensor
for the electromagnetic field. For a local medium response, we
introduce the permittivity, permeability and magnetoelectric
susceptibilities to obtain an explicitly duality-invariant scheme. We
find that duality invariance only holds as a continuous symmetry when
non-reciprocal responses are allowed for.
\end{abstract}

\maketitle


The linear response of a macroscopic material to applied
electromagnetic fields can go beyond the scope of simple descriptions
via electric permittivities and magnetic permeabilities
\cite{sihvolabook}. In particular, cross-suscepti\-bilities naturally
arise in chiral (meta-)materials \cite{chiral}, topological insulators
\cite{topological} or moving media \cite{moving}. An additional
complication arises in the latter case where the Onsager reciprocity
\cite{Onsager31} fails to hold. The Onsager principle is also
violated in Tellegen media \cite{Tellegen}, including the recently
proposed perfect electromagnetic conductor (PEMC) that continuously
interpolates between a perfect conductor and an
infinitely permeable material \cite{Sihvola}.

Chiral and anisotropic meta-materials have recently been discussed
as candidates for repulsive Casimir forces \cite{Soukoulis09}. Similar
effects have been predicted for topological insulators
\cite{topologicalCasimir} and materials with Chern--Simons interaction
\cite{Marachevskiy10}. Note that repulsive forces for magnetoelectric
media do not require chiral properties; they have originally been
discussed for dielectric plates interacting with magnetic ones
\cite{Henkel05}. To implement these effects with metamaterials, the
anisotropic response of the medium needs to be taken into account
\cite{DaRosa}.

The impact of electric and magnetic material properties can be studied
in a systematic way by means of a duality transformation
\cite{duality}. It has recently been shown that macroscopic QED in
isotropic magnetoelectrics obeys a discrete duality symmetry
\cite{Buhmann09}. This has immediate consequences for dispersion
forces in free space.

In this Letter, we develop the framework for studying quantum optical
phenomena in the presence of the most general linear absorbing media,
including nonlocal, bianisotropic and Onsager-violating materials. We
show that for locally responding media, the electromagnetic field can
be quantised in an explicitly duality-invariant way. The inclusion of
Onsager-violating materials restores duality as a continuous symmetry.
The successful quantisation is in contrast to recent claims that
canonical quantisation can only be performed for reciprocal media
\cite{Philbin10}. Our theory lays a solid foundation for studying
novel phenomena in the context of dispersion forces, Purcell effect
and quantum light propagation.


\paragraph*{Field quantisation for non-local media.}
The constitutive relation of a linear medium can be given by Ohm's
law in its most general form
\begin{multline}
\label{eq3}
\hat{\vect{j}}_\mathrm{in}(\vect{r},t)
=\int_{-\infty}^\infty\dif\tau\int\dif^3r'\,
\ten{Q}(\vect{r},\vect{r}',\tau)\sprod\hat{\vect{E}}(\vect{r}',t-\tau)
\\
+\hat{\vect{j}}_\mathrm{N}(\vect{r},t)\,.
\end{multline}
Here, $\ten{Q}(\vect{r},\vect{r}',\tau)$ is the conductivity tensor
and $\hat{\vect{j}}_\mathrm{N}(\vect{r},t)$ is the random noise
current required to fulfil the fluc\-tuation--dissipation
theorem~(\ref{eq21}) as given below. Causality requires that
$\ten{Q}(\vect{r},\vect{r}',\tau)\!=\!\tenszero$ for
$c\tau\!<\!|\vect{r}-\vect{r}'|$, in particular for all $\tau\!<\!0$.
Ohm's law in frequency space [$\hat{O}\!=\!\int_0^\infty\dif\omega\,%
\hat{\underline{O}}(\omega)+\mathrm{H.c.}$] takes the simple form
\begin{equation}
\label{eq6}
\hat{\underline{\vect{j}}}_\mathrm{in}(\vect{r},\omega)
=\int\dif^3r'\,\ten{Q}(\vect{r},\vect{r}',\omega)\sprod
 \hat{\underline{\vect{E}}}(\vect{r}',\omega)
+\hat{\underline{\vect{j}}}_\mathrm{N}(\vect{r},\omega)\,.
\end{equation}
with
\begin{equation}
\label{eq7}
\ten{Q}(\vect{r},\vect{r}',\omega)
=2\pi\underline{\ten{Q}}(\vect{r},\vect{r}',\omega)
=\int_0^\infty\dif\tau\,\me^{\mi\omega\tau}
\ten{Q}(\vect{r},\vect{r}',\tau)\;.
\end{equation}
As a result of the causality requirement, the conductivity obeys the
Schwarz reflection principle,
\begin{equation}
\label{eq8}
\ten{Q}^\ast(\vect{r},\vect{r}',\omega)
=\ten{Q}(\vect{r},\vect{r}',-\omega^\ast).
\end{equation}

Combining the constitutive relation with the Maxwell equations,
\begin{gather}
\label{eq4}
\vect{\nabla}\sprod\hat{\underline{\vect{E}}}
=\frac{\hat{\underline{\rho}}_\mathrm{\,in}}{\varepsilon_0}\,,\qquad
\vect{\nabla}\vprod\hat{\vect{\underline{E}}}
-\mi\omega\hat{\underline{\vect{B}}}=\veczero\;,\\
\label{eq5}
\vect{\nabla}\sprod\hat{\underline{\vect{B}}}=0,\qquad
\vect{\nabla}\vprod\hat{\underline{\vect{B}}}
+\frac{\mi\omega}{c^2}\,\hat{\underline{\vect{E}}}
 =\mu_0\hat{\underline{\vect{j}}}_\mathrm{in}\;,
\end{gather}
one finds that the electric field obeys a generalised inhomogeneous
Helmholtz equation
\begin{multline}
\label{eq9}
\biggl[\vect{\nabla}\vprod\vect{\nabla}\vprod
 \,\,-\,\frac{\omega^2}{c^2}\biggr]
\hat{\underline{\vect{E}}}(\vect{r},\omega)
-\mi\mu_0\omega\!\int\!\dif^3r'
 \ten{Q}(\vect{r},\vect{r}',\omega)\sprod
 \hat{\underline{\vect{E}}}(\vect{r}',\omega)\\
=\mi\mu_0\omega\hat{\underline{\vect{j}}}_\mathrm{N}(\vect{r},\omega).
\end{multline}
Introducing the Green tensor
\begin{multline}
\label{eq10}
\biggl[\vect{\nabla}\vprod\vect{\nabla}\vprod
 \,\,-\,\frac{\omega^2}{c^2}\biggr]
 \ten{G}(\vect{r},\vect{r}',\omega)\\
-\mi\mu_0\omega\!\int\!\dif^3s
 \ten{Q}(\vect{r},\vect{s},\omega)\sprod
 \ten{G}(\vect{s},\vect{r}',\omega)
=\bm{\delta}(\vect{r}-\vect{r}')
\end{multline}
with $\ten{G}(\vect{r},\vect{r}',\omega)\!\to\!\tenszero$ for
$|\vect{r}-\vect{r}'|\!\to\!\infty$, the formal solution to the above
integro-differential equation reads 
\begin{equation}
\label{eq11}
\underline{\vect{E}}(\vect{r},\omega)
=\mi\mu_0\omega\!\int\!\dif^3r'
 \ten{G}(\vect{r},\vect{r}',\omega)\sprod
 \hat{\underline{\vect{j}}}_\mathrm{N}(\vect{r}',\omega).
\end{equation}

By virtue of its definition~(\ref{eq10}), the Green tensor inherits
the Schwarz reflection principle~(\ref{eq8}) from the conductivity,
\begin{equation}
\label{eq12}
\ten{G}^\ast(\vect{r},\vect{r}',\omega)
=\ten{G}(\vect{r},\vect{r}',-\omega^\ast).
\end{equation}

As a major departure from previous treatments, we do not require the
conductivity to obey reciprocity, i.e.
$\ten{Q}^\trans(\vect{r}',\vect{r},\omega)\!=\!%
\ten{Q}(\vect{r},\vect{r}',\omega)$ does not necessarily hold. As a
consequence, the Green tensor will not obey the Onsager principle,
i.e.,
\begin{equation}
\label{eq14}
\ten{G}^\trans(\vect{r}',\vect{r},\omega)
=\ten{G}(\vect{r},\vect{r}',\omega)
\end{equation}
will not hold in general.

Despite this generalisation, it is still possible to derive an
integral relation for the Green tensor. To that end, we write the
Helmholtz equation~(\ref{eq10}) in the form
$\hat{\ten{H}}\sprod\hat{\ten{G}}\!=\!\hat{\ten{I}}$ where
$\langle\vect{r}|\hat{\ten{G}}|\vect{r}'\rangle%
\!=\!\ten{G}(\vect{r},\vect{r}',\omega)$ and
$\langle\vect{r}|\hat{\ten{H}}|\vect{r}'\rangle%
\!=\![\vect{\nabla}\vprod\vect{\nabla}\vprod-\omega^2/c^2]%
\bm{\delta}(\vect{r}-\vect{r}')
-\mi\mu_0\omega\ten{Q}(\vect{r},\vect{r}',\omega)$. The Green
operator is the right-inverse and, within any group of invertible
operators, also the left-inverse of the Helmholtz
operator, $\hat{\ten{G}}\sprod\hat{\ten{H}}\!=\!\hat{\ten{I}}$. From
this relation and its Hermitian conjugate, we find
\begin{equation}
\label{eq15}
\hat{\ten{G}}\sprod\bigl(\hat{\ten{H}}-\hat{\ten{H}}{}^\dagger\bigr)
\sprod\hat{\ten{G}}{}^\dagger
=\hat{\ten{G}}{}^\dagger-\hat{\ten{G}}.
\end{equation}
In coordinate space, this relation reads
\begin{multline}
\label{eq16}
\mu_0\omega\int\dif^3s\int\dif^3s'\,
\ten{G}(\vect{r},\vect{s},\omega)\sprod
\GRe\ten{Q}(\vect{s},\vect{s}',\omega)\sprod
\ten{G}^{\dag}(\vect{r}',\vect{s}',\omega)\\
=\GIm\ten{G}(\vect{r},\vect{r}',\omega).
\end{multline}
Here, we have introduced generalised real and imaginary parts of a
tensor field according to
\begin{gather}
\label{eq17}
\GRe\ten{T}(\vect{r},\vect{r}')=\tfrac{1}{2}\bigl[
 \ten{T}(\vect{r},\vect{r}')
 +\ten{T}^\dagger(\vect{r}',\vect{r})\bigr],\\
\label{eq18}
\GIm\ten{T}(\vect{r},\vect{r}')=\tfrac{1}{2\mi}\bigl[
 \ten{T}(\vect{r},\vect{r}')
 -\ten{T}^\dagger(\vect{r}',\vect{r})\bigr].
\end{gather}
They reduce to ordinary real and imaginary parts for orthogonal
tensor fields with $\ten{T}^\trans(\vect{r}',\vect{r})\!=\!%
\ten{T}(\vect{r},\vect{r}')$. The integral relation~(\ref{eq16})
generalises the result from Ref.~\cite{Raabe07} to the case where
Onsager reciprocity does not hold.

Returning to the electromagnetic field, an explicit quantisation is
achieved by specifying the commutator
\begin{equation}
\label{eq19}
\Bigl[\hat{\underline{\vect{j}}}_\mathrm{N}(\vect{r},\omega),
\hat{\underline{\vect{j}}}{}_\mathrm{N}^\dagger
 (\vect{r}',\omega')\Bigr]
=\frac{\hbar\omega}{\pi}\,\GRe\ten{Q}(\vect{r},\vect{r}',\omega)
 \delta(\omega-\omega').
\end{equation}
The fact that the right-hand side is a Hermitian tensor field
guarantees the consistency of this commutation relation. Introducing
the ground state $|\{0\}\rangle$ of the medium-field system according
to $\hat{\underline{\vect{j}}}_\mathrm{N}(\vect{r},\omega)|\{0\}%
\rangle\!=\!\veczero$, the currents satisfy the
fluctuation--dissipation theorem
\begin{multline}
\label{eq21}
\bigl\langle\bigl\{
 \Delta\hat{\vect{j}}_\mathrm{N}(\vect{r},\omega),
 \Delta\hat{\vect{j}}^\dagger_\mathrm{N}(\vect{r}',\omega')\bigr\}
\bigr\rangle\\
=\frac{\hbar}{\pi}\,\GIm[\mi\omega
 \ten{Q}(\vect{r},\vect{r}',\omega)]\delta(\omega-\omega')\;.
\end{multline}
Combining Eqs.~(\ref{eq11}) and (\ref{eq19}), one finds that the
fluctuations of the electric field are also consistent with the
fluctuation--dissipation theorem, as required:
\begin{multline}
\label{eq22}
\bigl\langle\bigl\{\Delta\hat{\vect{E}}(\vect{r},\omega),
 \Delta\hat{\vect{E}}^\dagger(\vect{r}',\omega')
 \bigr\}\bigr\rangle\\
=\frac{\hbar}{\pi}\,\GIm\bigl[\mu_0\omega^2
 \ten{G}(\vect{r},\vect{r}',\omega)\bigr]
 \delta(\omega-\omega').
\end{multline}

To verify the canonical equal-time commutation relations, we introduce
the vector potential for the electromagnetic field in the Coulomb
gauge, $\hat{\underline{\vect{A}}}\!=\!%
\hat{\underline{\vect{E}}}{}^\perp/(\mi\omega)$ ($\perp$: transverse
part). Using Eqs.~(\ref{eq11}) and (\ref{eq19}), one finds
\begin{multline}
\label{eq23}
 \bigl[\underline{\hat{\vect{E}}}(\vect{r},\omega),
 \underline{\hat{\vect{A}}}{}^\dagger(\vect{r}',\omega')\bigr]\\
=\frac{\mi\hbar\mu_0\omega}{\pi}
 \,\GIm\ten{G}^\perp(\vect{r},\vect{r}',\omega)
 \delta(\omega-\omega')
\end{multline}
and hence
\begin{multline}
\label{eq24}
 \bigl[\hat{\vect{E}}(\vect{r}),\hat{\vect{A}}(\vect{r}')\bigr]\\
=\frac{\hbar\mu_0}{2\pi}
 \int_{-\infty}^\infty\dif\omega\,\omega\,
 \bigl[\ten{G}^\perp(\vect{r},\vect{r}',\omega)
 +{}^\perp\ten{G}^\trans(\vect{r}',\vect{r},\omega)\bigr]
\end{multline}
where the Schwarz reflection principle~(\ref{eq12}) has been used.
Closing the integration contour in the upper half of the complex
$\omega$ plane and using the asymptote $(\omega^2/c^2)%
\ten{G}(\vect{r},\vect{r}',\omega)\!\to\!%
-\bm{\delta}(\vect{r}-\vect{r}')$ for $\omega\!\to\!\infty$,
one finds the canonical commutation relation from free-space QED,
\begin{equation}
\label{eq25}
 \bigl[\hat{\vect{E}}(\vect{r}),\hat{\vect{A}}(\vect{r}')\bigr]\\
=\frac{\mi\hbar}{\varepsilon_0}\,
\bm{\delta}^\perp(\vect{r}-\vect{r}').
\end{equation}

Next, we introduce bosonic creation and annihilation operators
$\hat{\vect{f}}{}^\dagger$, $\hat{\vect{f}}$ according to the
prescription
\begin{equation}
\label{eq26}
\hat{\vect{j}}_\mathrm{N}(\vect{r},\omega)
 =\sqrt{\frac{\hbar\omega}{\pi}}\int\dif^3r'\,
 \ten{R}(\vect{r},\vect{r}',\omega) \cdot
\hat{\vect{f}}(\vect{r}',\omega)
\end{equation}
where $\ten{R}$ is a square root of the positive definite tensor
field $\GRe\ten{Q}$,
\begin{equation}
\label{eq27}
\int\dif^3s\,\ten{R}(\vect{r},\vect{s},\omega)\sprod
 \ten{R}^\dagger(\vect{r}',\vect{s},\omega)
 =\GRe\ten{Q}(\vect{r},\vect{r}',\omega).
\end{equation}
Together with Eq.~(\ref{eq19}), this ensures bosonic
commutation relations,
$[\hat{\vect{f}}(\vect{r},\omega),%
\hat{\vect{f}}^\dagger(\vect{r}',\omega')]%
\!=\!\bm{\delta}(\vect{r}-\vect{r}')\delta(\omega-\omega')$.
The Hamiltonian of the body-field system is then
\begin{equation}
\label{eq29}
\hat{H}_\mathrm{F}
 =\int\dif^3r \int_0^\infty\dif\omega\,\hbar\omega\,
 \hat{\vect{f}}^\dagger(\vect{r},\omega)
 \sprod\hat{\vect{f}}(\vect{r},\omega).
\end{equation}
It leads to $\hat{\vect{f}}(\vect{r},\omega,t)\!=\!%
\hat{\vect{f}}(\vect{r},\omega)\me^{-\mi\omega t}$, hence Maxwell's
equations for the electromagnetic-field operators in the
Heisenberg picture are valid by construction.


\paragraph*{Field quantisation for local bianisotropic media.}
For media with a spatially local response, it is convenient to cast
the inhomogeneous Maxwell equations~(\ref{eq5}) into the alternative
forms
\begin{equation}
\label{eq30}
\vect{\nabla}\sprod\underline{\hat{\vect{D}}}=0,\qquad
\vect{\nabla}\vprod\underline{\hat{\vect{H}}}
 +\mi\omega\underline{\hat{\vect{D}}}
 =\veczero
\end{equation}
with
\begin{equation}
\label{eq31}
\underline{\hat{\vect{D}}}=\varepsilon_0\underline{\hat{\vect{E}}}
 +\underline{\hat{\vect{P}}},\qquad
\underline{\hat{\vect{H}}}=\frac{1}{\mu_0}\,\underline{\hat{\vect{B}}}
-\underline{\hat{\vect{M}}}.
\end{equation}
The polarisation and magnetisation fields respond linearly to the
electric and magnetic fields ($Z_0=\sqrt{\mu_0/\varepsilon_0}\,$),
\begin{eqnarray}
\label{eq32}
\underline{\hat{\vect{P}}}
&=&\varepsilon_0(\bm{\varepsilon}-\bm{\xi}\sprod\bm{\mu}^{-1}\sprod
 \bm{\zeta}-\ten{I})\sprod\underline{\hat{\vect{E}}}
+Z_0^{-1}\bm{\xi}\sprod\bm{\mu}^{-1}\sprod\underline{\hat{\vect{B}}}
 +\underline{\hat{\vect{P}}}_\mathrm{N},\nonumber\\ \\
\label{eq33}
\underline{\hat{\vect{M}}}
&=&Z_0^{-1}\bm{\mu}^{-1}\sprod\bm{\zeta}
 \sprod\underline{\hat{\vect{E}}}
+\mu_0^{-1}(\ten{I}-\bm{\mu}^{-1})\sprod\underline{\hat{\vect{B}}}
 +\underline{\hat{\vect{M}}}_\mathrm{N},
\end{eqnarray}
where $\bm{\varepsilon}(\vect{r},\omega)$ is the medium's
permittivity; $\bm{\mu}(\vect{r},\omega)$ its permeability;
$\bm{\xi}(\vect{r},\omega)$ and $\bm{\zeta}(\vect{r},\omega)$ its
magnetoelectric susceptibilities;
$\underline{\hat{\vect{P}}}_\mathrm{N}$ and
$\underline{\hat{\vect{M}}}_\mathrm{N}$ denote the noise polarisation
and magnetisation. Combining Eqs.~(\ref{eq31})--(\ref{eq33}), the
constitutive relations can be given in the more familiar form
\begin{eqnarray}
\label{eq34}
\underline{\hat{\vect{D}}}
&=&\varepsilon_0\bm{\varepsilon}\sprod\underline{\hat{\vect{E}}}
+c^{-1}\bm{\xi}\sprod\underline{\hat{\vect{H}}}
 +\underline{\hat{\vect{P}}}_\mathrm{N}
+c^{-1}\bm{\xi}\sprod\underline{\hat{\vect{M}}}_\mathrm{N},\\
\label{eq35}
\underline{\hat{\vect{B}}}
&=&c^{-1}\bm{\zeta}\sprod\underline{\hat{\vect{E}}}
+\mu_0\bm{\mu}\sprod\underline{\hat{\vect{H}}}
 +\mu_0\bm{\mu}\sprod\underline{\hat{\vect{M}}}_\mathrm{N}.
\end{eqnarray}

Combining Eqs.~(\ref{eq30}), (\ref{eq34}) and (\ref{eq35}), we see
that they are a special case of Eq.~(\ref{eq9}) with
\begin{align}
\label{eq36}
&\ten{Q}(\vect{r},\vect{r}',\omega)
=(\mi\mu_0\omega)^{-1}\vect{\nabla}\vprod(\bm{\mu}^{-1}\!-\!\ten{I})
\sprod\bm{\delta}(\vect{r}\!-\!\vect{r}')
\vprod\overleftarrow{\vect{\nabla}}{}'\nonumber\\
&\quad+Z_0^{-1}\vect{\nabla}\vprod\bm{\mu}^{-1}\sprod\bm{\zeta}\sprod
 \bm{\delta}(\vect{r}\!-\!\vect{r}')
+Z_0^{-1}\bm{\xi}\sprod\bm{\mu}^{-1}\sprod
 \bm{\delta}(\vect{r}\!-\!\vect{r}')
 \vprod\overleftarrow{\vect{\nabla}}{}'\nonumber\\
&\quad-\mi\varepsilon_0\omega
 (\bm{\varepsilon}\!-\!\bm{\xi}\sprod\bm{\mu}^{-1}
 \sprod\bm{\zeta}\!-\!\ten{I})
 \sprod\bm{\delta}(\vect{r}\!-\!\vect{r}'),\\
\label{eq37}
&\hat{\underline{\vect{j}}}_\mathrm{N}
 =-\mi\omega\hat{\underline{\vect{P}}}_\mathrm{N}
 +\vect{\nabla}\vprod\hat{\underline{\vect{M}}}_\mathrm{N}.
\end{align}
The Green tensor for the electric field (\ref{eq11}) solves
\begin{multline}
\label{eq38}
\biggl[\vect{\nabla}\vprod\bm{\mu}^{-1}\sprod\vect{\nabla}\vprod\,\,
-\,\frac{\mi\omega}{c}\vect{\nabla}\vprod\bm{\mu}^{-1}\sprod
 \bm{\zeta}\sprod\,\,
+\,\frac{\mi\omega}{c}\bm{\xi}\sprod\bm{\mu}^{-1}\sprod
 \vect{\nabla}\vprod\\
-\,\frac{\omega^2}{c^2}
 (\bm{\varepsilon}\!-\!\bm{\xi}\sprod\bm{\mu}^{-1}
 \sprod\bm{\zeta})\sprod\biggr]
 \ten{G}(\vect{r},\vect{r}',\omega)
 =\bm{\delta}(\vect{r}\!-\!\vect{r}').
\end{multline}

The commutation relations for $\hat{\vect{P}}_\mathrm{N}$ and
$\hat{\vect{M}}_\mathrm{N}$ can be deduced by substituting
Eqs.~(\ref{eq36}) and (\ref{eq37}) into Eq.~(\ref{eq19}),
\begin{align}
\label{eq39}
&\bigl[\hat{\underline{\vect{P}}}_\mathrm{N}(\vect{r},\omega),
\hat{\underline{\vect{P}}}{}_\mathrm{N}^\dagger
 (\vect{r}',\omega')\bigr]\nonumber\\
&\quad=\frac{\varepsilon_0\hbar}{\pi}\,\GIm
 (\bm{\varepsilon}\!-\!\bm{\xi}\sprod\bm{\mu}^{-1}
 \sprod\bm{\zeta})\delta(\vect{r}-\vect{r}')
 \delta(\omega-\omega'),\\
\label{eq40}
&\bigl[\hat{\underline{\vect{P}}}_\mathrm{N}(\vect{r},\omega),
\hat{\underline{\vect{M}}}{}_\mathrm{N}^\dagger
 (\vect{r}',\omega')\bigr]\nonumber\\
&\quad=\frac{\hbar}{2\pi\mi Z_0}\,
 (\bm{\zeta}^\dagger\sprod\bm{\mu}^{-1\dagger}
 \!-\!\bm{\xi}\sprod\bm{\mu}^{-1})
 \delta(\vect{r}-\vect{r}')\delta(\omega-\omega'),\\
\label{eq42}
&\bigl[\hat{\underline{\vect{M}}}_\mathrm{N}(\vect{r},\omega),
\hat{\underline{\vect{M}}}{}_\mathrm{N}^\dagger
 (\vect{r}',\omega')\bigr]\nonumber\\
&\quad=-\frac{\hbar}{\pi\mu_0}\,\GIm
 \bm{\mu}^{-1}\delta(\vect{r}-\vect{r}')
 \delta(\omega-\omega').
\end{align}
We introduce bosonic creation and annihilation operators
$[\hat{\vect{f}}_\lambda(\vect{r},\omega),%
\hat{\vect{f}}_{\lambda'}^\dagger(\vect{r}',\omega')]%
\!=\!\delta_{\lambda\lambda'}\bm{\delta}(\vect{r}-\vect{r}')%
\delta(\omega-\omega')$ ($\lambda,\lambda'\!=\!e,m$) according to
\begin{equation}
\label{eq44}
\begin{pmatrix}\hat{\vect{P}}_\mathrm{N}(\vect{r},\omega)\\
\hat{\vect{M}}_\mathrm{N}(\vect{r},\omega)\end{pmatrix}
 =\sqrt{\frac{\hbar}{\pi}}\,\mathcal{R}\sprod
 \begin{pmatrix}\hat{\vect{f}}_e(\vect{r},\omega)\\
 \hat{\vect{f}}_m(\vect{r},\omega)\end{pmatrix}\,,
\end{equation}
where the $(6\times 6)$-matrix $\mathcal{R}$ is a root of
\begin{equation}
\label{eq45}
\mathcal{R}\sprod\mathcal{R}^\dagger
=\begin{pmatrix}\varepsilon_0\GIm
 (\bm{\varepsilon}\!-\!\bm{\xi}\sprod\bm{\mu}^{-1}\sprod\bm{\zeta})
&\displaystyle\frac{\bm{\zeta}^\dagger\sprod\bm{\mu}^{-1\dagger}
 \!-\!\bm{\xi}\sprod\bm{\mu}^{-1}}{2\mi Z_0}\\
\displaystyle-\frac{\bm{\mu}^{-1}\sprod\bm{\zeta}
 \!-\!\bm{\mu}^{-1\dagger}\sprod\bm{\xi}^\dagger}{2\mi Z_0}
&\displaystyle-\frac{\GIm\bm{\mu}^{-1}}{\mu_0}
 \end{pmatrix}.
\end{equation}
The Hamiltonian of the body--field system is again quadratic and
diagonal in the bosonic variables,
\begin{equation}
\label{eq46}
\hat{H}_\mathrm{F}
 =\sum_{\lambda=e,m}\int\dif^3r \int_0^\infty\dif\omega\,\hbar\omega\,
 \hat{\vect{f}}_\lambda^\dagger(\vect{r},\omega)
 \sprod\hat{\vect{f}}_\lambda(\vect{r},\omega).
\end{equation}

Note that Eqs.~(\ref{eq31})--(\ref{eq33}) imply a separation of the
internal current density into electric and magnetic parts, 
$\hat{\underline{\vect{j}}}_\mathrm{in}
\!=\!-\mi\omega\hat{\underline{\vect{P}}}
\!+\!\vect{\nabla}\vprod\hat{\underline{\vect{M}}}$. This separation
and the resulting explicit field quantisation is not unique
\cite{Melrose}.


\paragraph*{Duality invariance.}
Introducing dual-pair notation
$(\hat{\vect{E}}^\trans,Z_0\hat{\vect{H}}^\trans)^\trans$,
$(Z_0\hat{\vect{D}}^\trans,\hat{\vect{B}}^\trans)^\trans$, we may
write the Maxwell equations (\ref{eq4}) and (\ref{eq30}) in the
compact form
\begin{gather}
\label{eq47}
\vect{\nabla}\sprod
 \biggl(\begin{array}{c}Z_0\hat{\underline{\vect{D}}}\\
 \hat{\underline{\vect{B}}}\end{array}\biggr)
=\biggl(\begin{array}{c}0\\ 0\end{array}\biggr),\\
\label{eq48}
\vect{\nabla}\vprod
 \biggl(\begin{array}{c}\hat{\underline{\vect{E}}}\\
 Z_0\hat{\underline{\vect{H}}}\end{array}\biggr)
 -\mi\omega
 \biggl(\begin{array}{cc}0&1\\-1&0\end{array}\biggr)
 \biggl(\begin{array}{c}Z_0\hat{\underline{\vect{D}}}\\
 \hat{\underline{\vect{B}}}\end{array}\biggr)
 =\biggl(\begin{array}{c}\veczero\\ \veczero\end{array}\biggr).
\end{gather}
The constitutive relations~(\ref{eq34}) and (\ref{eq35}) read
\begin{equation}
\label{eq49}
 \biggl(\begin{array}{c}Z_0\hat{\underline{\vect{D}}}\\
 \hat{\underline{\vect{B}}}\end{array}\biggr)
 =\frac{1}{c}\biggl(\begin{array}{cc}\bm{\varepsilon}&\bm{\xi}\\
 \bm{\zeta}&\bm{\mu}\end{array}\biggr)
 \biggl(\begin{array}{c}\hat{\underline{\vect{E}}}\\
 Z_0\hat{\underline{\vect{H}}}\end{array}\biggr)
 +\biggl(\begin{array}{cc}1&\bm{\xi}\\0&\bm{\mu}\end{array}\biggr)
 \biggl(\begin{array}{c}Z_0\hat{\underline{\vect{P}}}_\mathrm{N}\\
 \mu_0\hat{\underline{\vect{M}}}_\mathrm{N}\end{array}\biggr).
\end{equation}

Maxwell's equations are invariant under duality transformations
\begin{equation}
\label{eq50}
\begin{pmatrix}\vect{x}\\ \vect{y}\end{pmatrix}^\dual
 =D(\theta)
\begin{pmatrix}\vect{x}\\ \vect{y}\end{pmatrix},
 \qquad D(\theta)
 =\begin{pmatrix}\cos\theta&\sin\theta\\
 -\sin\theta&\cos\theta\end{pmatrix},
\end{equation}
because the symplectic matrix in Eq.~(\ref{eq48}) commutes with
$D(\theta)$. From the constitutive relations, we find
transformed medium response functions
$(\bm{\varepsilon},\bm{\xi},\bm{\zeta},\bm{\mu})^{\trans\dual}
\!=\!\mathcal{D}(\theta)(\bm{\varepsilon},\bm{\xi},\bm{\zeta},%
\bm{\mu})^{\trans}$ with
\begin{align}
\label{eq50b}
&\mathcal{D}(\theta)=\\
&\begin{pmatrix}
 \cos^2\theta&\sin\theta\cos\theta&\sin\theta\cos\theta
 &\sin^2\theta\\
 -\sin\theta\cos\theta&\cos^2\theta&-\sin^2\theta
 &\sin\theta\cos\theta\\
 -\sin\theta\cos\theta&-\sin^2\theta&\cos^2\theta
 &\sin\theta\cos\theta\\
 \sin^2\theta&-\sin\theta\cos\theta&-\sin\theta\cos\theta
 &\cos^2\theta\\
 \end{pmatrix}\nonumber
\end{align}
as well as
\begin{equation}
\label{eq50c}
\biggl(\begin{array}{cc}1&\bm{\xi}\\0&\bm{\mu}
 \end{array}\biggr)^\dual
 \biggl(\begin{array}{c}Z_0\hat{\underline{\vect{P}}}_\mathrm{N}\\
 \mu_0\hat{\underline{\vect{M}}}_\mathrm{N}\end{array}\biggr)^\dual
=D(\theta)\biggl(\begin{array}{cc}1&\bm{\xi}\\0&\bm{\mu}
 \end{array}\biggr)
 \biggl(\begin{array}{c}Z_0\hat{\underline{\vect{P}}}_\mathrm{N}\\
 \mu_0\hat{\underline{\vect{M}}}_\mathrm{N}\end{array}\biggr).
\end{equation}

It is worth discussing a few special cases of bianisotropic media,
their characteristic features and behaviour under duality
transformations:
\begin{itemize}
\item \textit{Isotropic media}
($\bm{\varepsilon}\!=\!\varepsilon\ten{I}$, $\bm{\mu}\!=\!\mu\ten{I}$,
$\bm{\xi}\!=\!\bm{\zeta}\!=\!\tenszero$): Onsager
reciprocity~(\ref{eq14}) holds;
$\hat{\vect{P}}_\mathrm{N}$ and $\hat{\vect{M}}{}^\dagger_\mathrm{N}$
commute; generalised real and imaginary parts reduce to ordinary
ones; discrete duality symmetry.
\item \textit{Biisotropic media}
($\bm{\varepsilon}\!=\!\varepsilon\ten{I}$, $\bm{\mu}\!=\!\mu\ten{I}$,
$\bm{\xi}\!=\!\xi\ten{I}$, $\bm{\zeta}\!=\!\zeta\ten{I}$): generalised
real and imaginary parts in Eqs.~(\ref{eq39}) and (\ref{eq42}) reduce
to ordinary ones; continuous duality symmetry.
\item \textit{Anisotropic media}
($\bm{\xi}\!=\bm{\zeta}\!=\!\tenszero$):
$\hat{\vect{P}}_\mathrm{N}$ and $\hat{\vect{M}}{}^\dagger_\mathrm{N}$
commute; discrete duality symmetry.
\item \textit{Reciprocal media}
($\bm{\varepsilon}^\trans\!=\!\bm{\varepsilon}$,
$\bm{\xi}^\trans\!=\!-\bm{\zeta}$,
$\bm{\mu}^\trans\!=\!\bm{\mu}$): Eq.~(\ref{eq14}) holds;
generalised real and imaginary parts reduce to ordinary ones;
discrete duality symmetry.
\end{itemize}
Here, discrete duality symmetry means that the rotation angle is
restricted to values $\theta\!=\!n\pi/2$ with $n\!\in\!\mathbb{Z}$.

To distinguish reciprocal magnetoelectric susceptibilities from
non-reciprocal ones, one commonly writes
$\bm{\xi}\!=\!\bm{\chi}^\trans\!-\!\mi\bm{\kappa}^\trans$ and
$\bm{\zeta}\!=\!\bm{\chi}\!+\!\mi\bm{\kappa}$. Here, the chirality
tensor $\bm{\kappa}\!=\!(\bm{\zeta}\!-\!\bm{\xi}^\trans)/(2\mi)$
represents the reciprocal magnetoelectric response; and the
non-reciprocal magnetoelectric tensor 
$\bm{\chi}\!=\!(\bm{\zeta}\!+\!\bm{\xi}^\trans)/2$ vanishes for a
reciprocal medium.

To derive transformation laws for the Green tensor, we combine
Eqs.~(\ref{eq4}), (\ref{eq9}), (\ref{eq31}), (\ref{eq33}), and
(\ref{eq37}) to write
\begin{gather}
\label{eq51}
\biggl(\begin{array}{c}\hat{\underline{\vect{E}}}\\
 Z_0\hat{\underline{\vect{H}}}\end{array}\biggr)
=-c\mathcal{G}\sprod
\biggl(\begin{array}{c}Z_0\hat{\underline{\vect{P}}}_\mathrm{N}\\
 \mu_0\hat{\underline{\vect{M}}}_\mathrm{N}\end{array}\biggr),\\
\label{eq52}
\mathcal{G}=\biggl(\begin{array}{cc}\ten{G}_{ee}&\ten{G}_{em}\\
\bm{\mu}^{-1}\sprod(\ten{G}_{me}\!-\!\bm{\xi}\sprod\ten{G}_{ee})&
\bm{\mu}^{-1}\sprod(\ten{G}_{mm}\!-\!\bm{\xi}\sprod\ten{G}_{em})
+\bm{\delta}\end{array}\biggr)
\end{gather}
where integration over the second position variable is implied and we
have introduced the shorthand notations
$\ten{G}_{ee}\!=\!(\mi\omega/c)\ten{G}(\mi\omega/c)$,
$\ten{G}_{em}\!=\!(\mi\omega/c)\ten{G}\vprod%
\overleftarrow{\vect{\nabla}}'$,
$\ten{G}_{me}\!=\!\vect{\nabla}\vprod\ten{G}(\mi\omega/c)$, and
$\ten{G}_{mm}\!=\!\vect{\nabla}\vprod\ten{G}\vprod
\overleftarrow{\vect{\nabla}}'$. The transformed tensors follow upon
applying duality transformations on both sides of the equation. In
general, the result is very complex due to the presence of the medium
response functions in Eqs.~(\ref{eq50c}) and (\ref{eq52}). In the
special case of $\vect{r}$ and $\vect{r}'$ being in free space, one
obtains the simple result
$(\ten{G}_{ee},\ten{G}_{em},\ten{G}_{me},\ten{G}_{mm}\!+\!%
\bm{\delta})^{\trans\dual}\!=\!\mathcal{D}(\theta)%
(\ten{G}_{ee},\ten{G}_{em},\ten{G}_{me},\ten{G}_{mm}\!+\!%
\bm{\delta})^{\trans}$. The Green tensors then transform like the
medium response functions with Eq.~(\ref{eq50b}).


\paragraph*{Conclusions.}
Starting from the general Ohm's law, we have quantised the
electromagnetic field in the presence of non-local, non-reciprocal
media which satisfies (i) the canonical commutation relations from
free-space QED; (ii) the linear fluctuation--dissipation theorem; and
(iii) the macroscopic Maxwell equations. Key feature of the scheme is
the symmetrisation of tensor fields via generalised real and imaginary
parts, which is necessary whenever Onsager reciprocity does not hold.

For a local bianisotropic medium, we have shown that quantisation can
alternatively performed by introducing permittivity, permeability and
magnetoelectric susceptibilities. When the latter do not vanish,
the noise polarisation and magnetisation do not commute. We have
explicitly determined the behaviour of the fields, response functions
and Green tensor under duality transformations. The full continuous
transformation group applies for bianisotropic and biisotropic media,
but reduces to a discrete symmetry for anisotropic, isotropic and/or
reciprocal media.

The scheme lays the foundation for exact studies of quantum phenomena
such as dispersion forces, (F\"orster) energy transfer or
environment-assisted molecular transition rates in the presence of
motion or novel media with chiral or non-reciprocal properties. For
example, it will facilitate the design of materials that show Casimir
repulsion, and guide towards new schemes to test CP-violation in
atoms.

This work was supported by the UK Engineering and Physical Sciences
Research Council (EPSRC). We would like to thank F.~H. Hehl for
discussions.


\end{document}